\documentclass[conference]{IEEEtran}
\IEEEoverridecommandlockouts
\usepackage{pbalance}

\usepackage{cite}
\usepackage{amsmath,amssymb,amsfonts}
\usepackage{algorithmic}
\usepackage{graphicx}
\usepackage{wrapfig}
\usepackage{textcomp}
\usepackage{xcolor}
\usepackage{url}
\usepackage[most]{tcolorbox}
\def\BibTeX{{\rm B\kern-.05em{\sc i\kern-.025em b}\kern-.08em
    T\kern-.1667em\lower.7ex\hbox{E}\kern-.125emX}}

\begin{document}

\title{Towards a Definition of Complex Software System}

\author{
\IEEEauthorblockN{Jan Žižka}
\IEEEauthorblockA{0009-0007-6483-0037\\
Faculty of Informatics \\
Masaryk University \\
Brno, Czech Republic \\
Botanická 68a, Brno, 60200 \\
Email: jzi@mail.muni.cz}
\and
\IEEEauthorblockN{Bruno Rossi}
\IEEEauthorblockA{0000-0002-8659-1520\\
Faculty of Informatics \\
Masaryk University \\
Brno, Czech Republic \\
Botanická 68a, Brno, 60200 \\
Email: brossi@mail.muni.cz}
\and
\IEEEauthorblockN{Tomáš Pitner}
\IEEEauthorblockA{0000-0002-2933-2290\\
Faculty of Informatics \\
Masaryk University \\
Brno, Czech Republic \\
Botanická 68a, Brno, 60200 \\
Email: pitner@muni.cz}
}

\maketitle

\begin{abstract}
Complex Systems were identified and studied in different fields, such as physics, biology, and economics.
These systems exhibit exciting properties such as self-organization, robust order, and emergence.
In recent years, software systems displaying behaviors associated with Complex Systems are starting to appear, and these behaviors are showing previously unknown potential (e.g., GPT-based applications).
Yet, there is no commonly shared definition of a Complex Software System that can serve as a key reference for academia to support research in the area.
In this paper, we adopt the theory-to-research strategy to extract properties of Complex Systems from research in other fields, mapping them to software systems to create a formal definition of a Complex Software System. We support the evolution of the properties through future validation, and we provide examples of the application of the definition. Overall, the definition will allow for a more precise, consistent, and rigorous frame of reference for conducting scientific research on software systems. 
% 156 words
\end{abstract}


\section{Introduction}

\IEEEPARstart{C}{omplex} Systems manifest multifaceted dependencies and interrelationships with other systems and environments, making them difficult, if not impossible, to model in their entirety~\cite{thurner2018introduction, Ladyman2013}. Complex Systems show properties that make them peculiar, such as the property of \textit{emergent behavior}, i.e., behavior deriving from the different parts of a system that cannot easily be determined or forecasted when components are observed in isolation. 

Complex Systems theory focuses on understanding and explaining the behavior of Complex Systems formed by interacting components~\cite{thurner2018introduction}. The theory provides a general framework and a set of methodologies to study the emergent properties and dynamics embedded in Complex Systems. However, there is no agreed precise definition of the term as different authors might have other points of view~\cite{naturedisputeddefs}. Complex Systems have been studied in various fields \cite{mitchell2009complexity,klir1991architecture,waldrop1993complexity}, for example, in social sciences by exploring the complex interactions of individuals in cities.

The successes of software systems in the past years based on, for example, Neural Networks, such as systems developed by DeepMind \cite{deepmind}, or OpenAI \cite{openai}, bring a new range of software systems to wide attention. The appearance of applications such as AlphaFold or ChatGPT and others in a very short time suggests that many more such systems will rise soon.

The exciting behaviors of these new software systems, such as self-organization and emergence, cannot be explained by inspecting the software implementation they are based on. This range of software systems has specific behaviors correlating with the behaviors of Complex Systems as defined by the Complex Systems theory.

While in Computer Science, complexity is studied in different contexts, such as code complexity and the complexity of algorithms~\cite{van1991handbook}, this paper focuses on complexity in the context of \textit{software systems}. Also, many software systems are \textit{socio-technical} systems where humans are part of the system rather than only forming its environment. Our study is interested in pure technical systems where humans are not part of the system but may build the system's environment.

Our work aims to provide a clear definition of a Complex Software System (CSS) based on the \textit{theory-to-research} strategy \cite{swanson2013theory, reynolds2015primer}, providing a frame of reference about the properties of such systems in relation to what is postulated by Complex Systems' theory~\cite{thurner2018introduction}.

As the field of Complex Systems is evolving \cite{Ladyman2013}, we suggest a framework that will also allow the evolution of the definition and terms. The precise definitions will allow more straightforward and unambiguous communication within academia and will be able to connect to existing and future real-world Complex Software Systems. The definition will also provide boundaries for new research fields with a degree of focus cleared of possible ambiguity due to the lack of definitions.

To summarize, we have the following contribution in this article:
\begin{itemize}
    \item Setting up a framework under which the definition of a Complex Software System is created;
    \item Defining a Complex Software System based on reference to general Complex Systems theory;
    \item Listing examples of the use of such a definition;
    \item Based on the definition and proposed use, list potential future research directions;
\end{itemize}

The article is structured as follows. In Section~\ref{sec:basic-def}, we provide basic definitions that are commonly adopted in the context of software systems when discussing Complex Software Systems, such as System of Systems (SoS), Software Ecosystems (SECO), and Complex Adaptive Systems (CAS). The purpose of the need for a precise definition is discussed in Section \ref{why}.
Section~\ref{method} discusses the method for creating the definition of a Complex Software System.
We select several postulates in Section~\ref{definition} to form an initial base
for defining a Complex Software System. Section~\ref{examples} provides 
examples of using such a definition. Future research directions are discussed
in Section~\ref{future}, and conclusions are presented in Section \ref{conclusions}.

\section{Basic Definitions}
\label{sec:basic-def}

In Software Engineering, several commonly used terms and definitions of software systems discuss how software systems and components can be combined and aggregated. This section lists some of the main definitions and examines their relationship to Complex Software Systems.

\textbf{System of Systems (SoS)} is a collection of independent interacting systems~\cite{ackoff1971towards}. An SoS has several key properties~\cite{nielsen2015systems}:
\begin{itemize}
    \item Operational Independence. Any system part of an SoS is self-standing and can operate even if the whole SoS is disaggregated.
    \item Managerial Independence. Every single system in an SoS is self-governing.
    \item Geographic Distribution. SoS are often distributed over geographic regions.
    \item Evolutionary Development. The existence and development of SoS are under constant change.
    \item Emergent Behaviour. \emph{"Through the collaboration between the systems in an SoS, a synergism is reached in which the system behavior fulfills a purpose that cannot be achieved by, or attributed to, any of the individual systems."} \cite{nielsen2015systems}
\end{itemize}

The systems which are part of an SoS may also be Complex Systems, or the SoS as a whole may form a Complex System -- however, the definition of an SoS does not imply that such a system is a Complex System.  

\textbf{Software Ecosystems (SECO)}  are \textit{"defined as a set of businesses functioning as a unit and interacting with a shared market for software and services, together with relationships among them"} \cite{messerschmitt2003software}.
A SECO may be composed of Complex Software Systems and is a type of SoS. SECOs are typically socio-technical systems \cite{lima2015survey}, which exhibit Complex System behaviors. In a SECO, introducing new elements can potentially have disruptive effects. SECOs features \cite{joshua2013software,lettner2014case} for example contain and provide:
\begin{itemize}
    \item Inherited characteristics of natural ecosystems like predation, parasitism, mutualism, commensalism, symbiosis, and amensalism.
    \item Architectural concepts like interface stability, evolution management, security, and reliability.
    \item Open source development model.
    \item Platform for negotiating requirements aligning needs with solutions, components, and portfolios.
    \item Capability for process innovation.
    \item Controlled central part for the core of the technology.
\end{itemize}

\textbf{Complex Adaptive Systems (CAS)} \textit{"are systems that have a large number of components, often called agents, that interact and adapt or learn"}~\cite{holland1992complex,holland2006studying}. The field of CAS focuses on the adaptive behavior of Complex Systems. 

Software CAS refers to software systems that exhibit emergent behavior and self-organization, similar to Complex Adaptive Systems found in nature. These systems can adapt and evolve based on their interactions with the environment through feedback loops. They involve multiple interacting components or agents that collectively exhibit behavior that cannot be easily predicted from the behavior of individual components~\cite{holland1992complex}.

A subset of Software CAS are Software Self-Adaptive Systems (SSAS)  that focus on the ability of a software system to autonomously adapt and modify the behavior or configuration in response to changing conditions or requirements~\cite{de2013software,macias2013self}. SSAS have built-in mechanisms that monitor the system's state, analyze environmental changes, and take actions to maintain or improve system properties at runtime~\cite{de2013software}. 

Software CAS~\cite{holland1992complex,holland2006studying}:
\begin{itemize}
    \item Typically operate far from equilibrium.
    \item Undergo revisions and improvements.
    \item Do not conform to classic, equilibrium-based mathematical approaches.
    \item Continuously adapt through recombination of the building blocks. 
\end{itemize}

We summarize the main characteristics of SoS, SECO, and CAS in Table~\ref{tbl:definitions}.

\begin{table*}[!t]
% increase table row spacing, adjust to taste
\renewcommand{\arraystretch}{1.3}
\caption{SoS, SECO, CAS Concepts}
\label{tbl:definitions}
\centering
\begin{tabular}{|p{2.5cm}|p{4.5cm}|p{4.5cm}|p{4.5cm}|}
\hline
 & \textbf{System of Systems (SoS)} & \textbf{Software Ecosystems (SECO)} & \textbf{Complex Adaptive Systems (CAS)}\\
\hline
\textbf{Definition} & Collection of independent interacting systems~\cite{ackoff1971towards} & Collection of software components, applications, and services~\cite{messerschmitt2003software} & Collection of components (agents) that interact and evolve~~\cite{holland1992complex,holland2006studying}\\
\hline
\textbf{Focus} & Collection of interacting systems & Software and services relationships & Software agents interaction\\
\hline
\textbf{Emergent Behavior} & As systems get larger, emergent behavior is more probable~\cite{osmundson2008kr14} &  Limited emergence, the introduction of new elements might have disrupting effects & Emergence in terms of adaptive behavior and self-organization \\
\hline
\textbf{Examples} & Smart Grids Systems & Android Ecosystem & Robotic Swarm\\
\hline
\end{tabular}
\end{table*}

\section {Why there is a need for a definition of a Complex Software System?} \label{why}

We need the definition of a Complex Software System for several reasons. Below we discuss benefits, which are the motivators for the research presented in this article.

\textbf{Clarity and precision:} ensure that the meaning of the term Complex Software System is unambiguous.

\textbf{Consistency:} avoid that a software Complex System is defined differently by different researchers or in other contexts, and ensure that the term's meaning remains consistent over time.

\textbf{Rigor:} provide a framework for scientific research.
Scientific definitions are necessary for the development of clear, precise, and consistent
scientific concepts and for the advancement of scientific research.

Once the definition has been developed, it can be used in various ways.
For example for:

\textbf{Hypothesis testing:} 
support the development of hypotheses about
properties of a Complex Software System and its behaviors so that they can be tested through experiments and observations.
For example, empirical methods can be used to verify if a software system fulfills the necessary conditions for forming a Complex Software System.

\textbf{Classification:} allow classification or categorization of Complex Software Systems
based on their properties. For example, software systems
use different paradigms, such as Neural Networks or multi-agent architectures.
This can be used to create classifications based on system boundaries, technology, or the form of their implementation.

\textbf{Comparison:} a known Complex Software Systems can be
compared based on the definition with other systems to find similar or differing properties.
This can serve as grounds for expanding the definition or driving the creation of similar
software systems.

\textbf{Theory development:} develop new theories or models.

Definitions are essential for academic research, providing a clear and precise framework for developing hypotheses, conducting experiments, and developing theories. Definitions allow researchers to communicate effectively and to build upon each other's work.

\section{Methodology to Build the Definition} \label{method}

To define a Complex Software System, we adopt the \textit{theory-to-research} strategy, in which there is a continuous cycle between theory and empirical validation~\cite{swanson2013theory, reynolds2015primer}:

\begin{itemize}
    \item we extract from the literature (e.g.,~\cite{thurner2018introduction,Ladyman2013}) common properties of Complex Systems as have been studied in the different fields;
    \item as there is no full agreement on all properties in the context of the theory
of Complex Systems~\cite{naturedisputeddefs}, we discuss the most appropriate in the context of software systems, both according to our view of software systems implementation and deployment and with the aid of further related works \cite{holland1992complex,ottino2003complex, waldrop1993complexity, mitchell2009complexity};
    \item we map each of the properties to a set of identified \textit{necessary}, \textit{sufficient} and \textit{representative conditions} to the context of software systems;
    \item we provide an initial application of the definition to showcase the main benefits;
\end{itemize}

% \section{Framework for Complex Software Systems definition} \label{framework}

Ladyman~\cite{Ladyman2013} defines a Complex System based on reviewing attempts in the literature to characterize a Complex System
and compiles a set of necessary conditions to represent complexity.
In their work, authors provide conditions that are qualitative and which may not be sufficient for complexity, but they set a basis for a way how complexity can be defined. We suggest using the same method for defining a Complex Software System.

We base the method of writing a formal definition of a Complex Software System on a list of properties of three types that are \textit{necessary}, \textit{sufficient}, and \textit{representative} conditions written in natural language. 

\textbf{The set of properties is the definition of a Complex Software System.}

The properties will be assembled in the form of graphical frames followed by a commentary.
Different types of properties will be color-coded for clarity in the following way,
where \emph{"n"} is an ordered number and \emph{"keyword"} is a word abbreviating
the property:

\begin{tcolorbox}[colback=gray!5!white,colframe=orange!75!black,title=Property n - \emph{"keyword"} - Necessary]
       A property that is a necessary condition but not sufficient to define a Complex Software System.
       Any Complex Software System must be fulfilling all properties that are necessary conditions.
       But fulfilling all such conditions doesn't imply that the Software System is Complex Software System.
\end{tcolorbox}

\begin{tcolorbox}[colback=gray!5!white,colframe=lime!75!black,title=Property n - \emph{"keyword"} - Sufficient]
       A property that is the sufficient condition to define a Complex Software System.
       If a software system fulfils any single property that is sufficient condition then such
       a system is a Complex Software System.
\end{tcolorbox}

\begin{tcolorbox}[colback=gray!5!white,colframe=cyan!75!black,title=Property n - \emph{"keyword"} - Representative]
       A property describing a typical feature of a Complex System is a representative property.
       Such property is neither necessary nor sufficient but does describe a commonality among Complex Software Systems.
\end{tcolorbox}

The nomenclature allows referring specific property such as \colorbox{orange!75}{Pn-N}~\emph{"keyword"}
for a necessary condition property, \colorbox{lime!75}{Pn-S}~\emph{"keyword"} for sufficient condition
property and \colorbox{cyan!75}{Pn-R}~\emph{"keyword"} for a representative property.

The numbering of properties is sequential across all the types. The intention and expectation
is that the type of the property may change based on future validation and research and keeping
the numbering intact will allow for unambiguous referencing.

\section{Initial Definition of a Complex Software System} \label{definition}

A Complex Software System may be defined by a set of properties which may be viewed as \textit{necessary},  \textit{sufficient}, and \textit{representative} for a system to exhibit Complex System behaviors.
Such properties are generic for any Complex System and are also described in existing publications such as \cite{thurner2018introduction, Ladyman2013}.
In this section, we will summarize the basic properties of Complex Systems and put those in the context of software systems, creating a base for the definition of a Complex Software System.

\begin{tcolorbox}[colback=gray!5!white,colframe=orange!75!black,title=Property 1 - \emph{"components"} - Necessary]
       A Complex Software System is composed of many components.
\end{tcolorbox}

All definitions of systems complexity \cite{thurner2018introduction, Ladyman2013,mitchell2009complexity,waldrop1993complexity} require the system to have many components.
In the context of Complex Systems, the word \emph{"many"} refers to the qualitative
rather than quantitative nature of the term. It would, therefore, be incorrect to attempt to quantify it.
For example, a system composed of two Complex Systems with manifold interactions and fulfilling
other necessary properties is a Complex System, as well as a system composed of millions of components of
a similar type, maybe a Complex System.
This property comes directly from the definition of the word "system"~\cite{mw:system,oed:system}. Software systems are typically composed of components. This property is a necessary condition but not sufficient for a software system to exhibit complexity. In software, a component may describe different entities based on view or perspective. It can represent a code module, software package, process, or service. From the perspective of a Complex Software System, only a subset of such representations can serve as components in a Complex Software System as
they must possess further attributes discussed in the following paragraphs.

\begin{tcolorbox}[colback=gray!5!white,colframe=orange!75!black,title=Property 2 - \emph{"communication"} Necessary]
       The components of a Complex Software System have means of intercommunication.
\end{tcolorbox}

Communication is an essential condition for a Complex Software System.
As Ladyman~\cite{Ladyman2013} explains: \emph{"Without interaction, a system merely forms a “soup” of particles which necessarily are independent and have no means of forming patterns, of establishing order."}. Communication through messaging
shared data, and interfaces is fundamental in software systems. However, this property is not a sufficient condition for a Complex Software System, as many software systems communicate but lack other necessary properties.

\begin{tcolorbox}[colback=gray!5!white,colframe=cyan!75!black,title=Property 3 - \emph{"similarity"} - Representative]
       The components in a Complex Software System are similar.
\end{tcolorbox}

Based on Ladyman~\cite{Ladyman2013}: \emph{"For interactions to happen and for pattern and coherence to develop, the elements have to be not only many but also similar in nature."}.
From the software systems perspective, for example, a system based on front-end, business logic middle-ware, and back-end database components may not form a Complex Software System. This has fascinating implications for software systems, which may be considered complex. However, this condition is not sufficient to determine a Complex Software System. This property may be necessary, but such a statement cannot be demonstrated and proven with the current knowledge and it is a question if a Software System with dis-similar components
may still form a Complex Software System or if the system boundaries would exclude such dis-similar
components into system's environment rather than being part of the system itself. It also remains
to be defined what precisely \emph{similar} means in the context of software components.

\label{property4}
\begin{tcolorbox}[colback=gray!5!white,colframe=orange!75!black,title=Property 4 - \emph{"interaction-change"} - Necessary]
       The strength of components interactions in a Complex Software System are not static, but they change over time.
\end{tcolorbox}

\emph{"Most interactions are mediated through some sort of exchange process between nodes (components).
In that sense, interaction strength is often related to the quantity of objects exchanged."} \cite{thurner2018introduction}.
The interactions among components have to change over time for a Complex Software System to evolve and create a self-organized clustered structure~\cite{thurner2018introduction}. The resulting network topology contains information about the dynamics and formation of the nodes and links (Chapter 4.5) \cite{thurner2018introduction}.
This is a familiar property in software systems studies, for example, in the field of dynamic or adaptive networks~\cite{kuhn2011dynamic, sayed2014adaptive}.
This property is necessary for a Complex Software System from which self-organization and clustering emerge.

\begin{tcolorbox}[colback=gray!5!white,colframe=orange!75!black,title=Property 5 - \emph{"states"} - Necessary]
       Components of a Complex Software System are characterized by states.
\end{tcolorbox}

Complex Software Systems are systems that evolve. An algorithmic description of evolution (Chapter~5) \cite{thurner2018introduction} is based on the fact that the system has states, and the evolution forms a path through states from time $t$ to time $t + 1$. Therefore the existence of states is necessary to create a Complex Software System. The notion of states in software systems is among the basic concepts of any information systems~\cite{sommerville2016software}. However, the existence of
states is not a sufficient condition for Complex Software Systems.

\begin{tcolorbox}[colback=gray!5!white,colframe=cyan!75!black,title=Property 6 - \emph{"co-evolution"} - Representative]
       The intercommunication and states of components in a Complex Software System are not independent but co-evolve.
\end{tcolorbox}

As discussed in (Chapter 1.5)~\cite{Ladyman2013} \emph{"Complex systems are characterized by the fact that states and interactions are often not independent but evolve together by mutually influencing each other; states and interactions co-evolve."}.
The Complex Systems are characterized by co-evolutionary dynamics (Chapter 4.8) \cite{thurner2018introduction}. From a software systems perspective, this can be represented, for example, by adaptive network models \cite{sayed2014adaptive}, which are known to exhibit such co-evolutionary dynamics~\cite{antoniades2015co,farajtabar2015co}.

\begin{tcolorbox}[colback=gray!5!white,colframe=cyan!75!black,title=Property 7 - \emph{"context-dependence"} - Representative]
       A Complex Software System is context-dependent.
\end{tcolorbox}

As shown by Thurner~\cite{thurner2018introduction}, the Complex Systems are often represented by multi-layer networks and \emph{"... for any dynamic process happening on a given network layer, the other layers represent the 'context' ..."}. In other words, such context defines how components on different layers may be influenced. This property is typical for Complex Systems to co-evolve through context dependency. 

\begin{tcolorbox}[colback=gray!5!white,colframe=cyan!75!black,title=Property 8 - \emph{"algorithmicism"} - Representative]
       A Complex Software System is algorithmic.
\end{tcolorbox}

Based on Thurner~\cite{thurner2018introduction}, the \emph{"... (Complex Systems) algorithmic nature is a direct consequence of the discrete interactions between interaction networks and states."}. This fits software systems that are naturally algorithmic \cite{sommerville2016software}.

\begin{tcolorbox}[colback=gray!5!white,colframe=cyan!75!black,title=Property 9 - \emph{"path-dependence"} - Representative]
       A Complex Software System processes are path-dependent and non-ergodic.
\end{tcolorbox}

\emph{"The Complex Systems are typically governed by path-dependent processes."}
(Section 2.5)~\cite{thurner2018introduction}. The process, in a general theory of complex
systems, refers to stochastic processes~\cite{ross1995stochastic}. This further means that processes in complex
systems are inherently non-Markovian. It can also be shown that Complex Systems
are non-ergodic (for in-depth discussion, see~\cite{thurner2018introduction}).
From the software systems perspective, this means that software
system to exhibit such Complex System properties, they must change their
boundary conditions as the system evolves. 

\begin{tcolorbox}[colback=gray!5!white,colframe=orange!75!black,title=Property 10 - \emph{"disorder"} - Necessary]
       A Complex Software System is disordered and out-of-equilibrium.
\end{tcolorbox}

Ladyman~\cite{Ladyman2013} argues that \emph{"...complex
systems are precisely those whose order emerges from a disorder rather than being built into them."}. Also, it can be noted that Complex Systems are generally out-of-equilibrium \cite{thurner2018introduction}, which
drives interesting challenges to the concepts of entropy. Although it can
be shown~\cite{Ladyman2013,thurner2018introduction} that Complex Systems exhibit such properties, it is not obvious how to apply those to software systems.

\begin{tcolorbox}[colback=gray!5!white,colframe=orange!75!black,title=Property 11 - \emph{"robust-order"} - Necessary]
       A Complex Software System exhibits robust order.
\end{tcolorbox}

The concept of robust order is derived from system disorder. As shown by Ladyman~\cite{Ladyman2013} \emph{"... a system consisting of many similar components (elements) which are
communicating (interacting) in a disordered way has the potential of forming patterns
or structures"}. This refers to self-organization and emergence property.
From a software system perspective, this indicates that a Complex Software System
shall be composed of similar and at least initially
disordered components. This might seem to contradict with \colorbox{orange!75}{P10-N}~\emph{"disorder"} but 
\emph{"... although the elements continue to interact in a disordered way, the overall patterns and structures are preserved. A macroscopic level arises out of microscopic interaction, and it is stable"}~\cite{Ladyman2013} which
Ladyman defines it as a robust order and continues that \emph{"t(T)his kind of robust order is a further necessary condition for a system to be complex"}. Therefore disorder and robust order may co-exist.
One example of software systems showing such a property are Artificial Neural Networks (ANN),
where initially, input weights of neurons may be initialized with random values and, through
learning, such initial disorder forms patterns, structures, or clusters. Also, when examined
on a neuron level, ANN will still be disordered.

\begin{tcolorbox}[colback=gray!5!white,colframe=orange!75!black,title=Property 12 - \emph{"memory"} - Necessary]
       A Complex Software System has memory.
\end{tcolorbox}

From Holland~\cite{holland1992complex}: 
\emph{"A system remembers through the persistence of internal structure"}, Ladyman~\cite{Ladyman2013} infer that \emph{"Memory is a straightforward corollary of robust order."}.
And Thurner~\cite{thurner2018introduction} notes that the \emph{"Complex systems often have memory.
Information about the past can be stored in nodes (components), if they have a memory,
or in the network structure of the various layers."}
In such a sense, \emph{memory} refers to the internal self-organized structure of the system.
The difference between \emph{memory} and \emph{states} defined by \colorbox{orange!75}{P5-N}~\emph{"states"}
is that \emph{states} represent the system at a specific point in time, but they do not represent
history-dependent dynamics stored in the systems \emph{memory}.
This property might have various interpretations in software systems, such as
a path through imitation-learning \cite{vinyals2019grandmaster} or system
audit trails. This interesting property might also have yet unknown interpretations
in software systems.

\begin{tcolorbox}[colback=gray!5!white,colframe=lime!75!black,title=Property 13 - \emph{"SoS sufficiency"} - Sufficient]
       A System of Complex Software Systems is a Complex Software System.
\end{tcolorbox}

As an intuitive analogy to properties of a Complex Software System -- as in Ackoff~\cite{ackoff1971towards} -- it may be possible to show that a system of Complex Software Systems forms a
Complex Software System. 

This might have exciting implications in practice as once a Complex Software System is created and exists, a new Complex Software System may be formed by creating a system of such systems (SoS).

As the software does not require any material or physical manipulation and software systems can be created relatively quickly, this allows the possible rapid advancement of software-based systems exhibiting Complex System behaviors. 

\section{Application of the definition} \label{examples}

\subsection{Unambiguous communication within academia}

"Complex Software System" is a widely used term
 in academia and industry. It refers to a wide range of software systems and viewpoints with a generic notion of a system's complexity.
The definition presented in this paper attempts to provide a concrete
reference that can be utilized throughout academic discussions
to facilitate a common understanding of the term and properties of
such a system. Also, the proposed definition framework is intended
to extend and refine the definition to support further Complex Software Systems theory development.

The definition of a Complex Software System may be referenced as a whole, or specific properties may be the focus of empirical and other research when studying the properties of software systems.
Having a definition of a Complex Software System will bring clarity through academic discussions.

\subsection{Complex Software System categorization}\label{categorization}

Software systems are open systems \cite{bertalanffy1968general} with external interactions.
The boundary of the system defines what belongs to the system itself and what its surroundings are. The edge of the system may be used for categorization. Many software systems nowadays are socio-technical systems \cite{tomic2021bionic} where people are part of the system rather than creating the surroundings and interacting with the system only through the system boundary.

The software systems also interact with humans or are part of machine-to-machine interactions.
The software system boundaries can be used as one of the aspects of categorizing types of software systems. Most importantly, this categorization has an essential perspective from Complex Software Systems theory. Most of the socio-technical systems are Complex Systems \cite{tomic2021bionic}, and the involvement of humans fulfills the necessary conditions presented in Section~\ref{definition}.

For example, if we consider the Internet as a Complex Software System,
it can be viewed as a socio-technical system. In which case, it fulfills the \colorbox{orange!75}{P4-N} "interaction-change" property. The changes are done by human developers, companies, and communities, which interconnect services throughout the Internet. If the boundary of the Internet as a system excludes human actors, the \colorbox{orange!75}{P4-N} "interaction-change" might not hold. 

The presented properties applied to different boundaries of a software
systems can, in this way, provide mechanisms to create a categorization and demonstrate which boundary is allowing the creation of a Complex System and which is not, as they are not fulfilling the necessary conditions defined by the presented properties.

\subsection{Complex Software System modeling}

To dive into understanding Complex Software Systems, it will
be required to have a model to analyze the properties' effects, how the \textit{necessary} and \textit{sufficient} conditions may be fulfilled or violated, and how \textit{representative} properties may help define a Complex Software System.
This can be achieved, for example, by studying existing Complex Systems, as it is done in other fields. However, the challenge is that we might not have access to such systems and, based on the boundary categorization~(Section~\ref{categorization}), some categories of Complex Software Systems might not even exist, for example, pure technical Complex Software Systems, where humans are outside the system boundary.

The model can be designed and developed to study such Complex Software Systems based on the presented definition and specification of necessary conditions for such a system to exist.

The model will allow experiments to evaluate the assumptions placed by the properties of Complex Software Systems. Understanding underlying principles might show how such software systems may
be constructed.

With models based on the defined Complex System properties, it may be, for example, demonstrated that \colorbox{lime!75}{P13-S} "SoS sufficiency" is a \textit{sufficient} condition.

Creating models of different categories of Complex Software Systems is another research path we will follow.

\section{Future research} \label{future}

Creation of the definition of a Complex Software System and
a framework for describing the definition will open doors
for several research areas:

\begin{itemize}
    \item Search for and refining Complex Software System properties;
    \item Exploring categories of Complex Software Systems;
    \item Creation of Complex Software System models;
    \item Search for underlying principles of Complex Software Systems;
\end{itemize}

The framework discussed in Section~\ref{method} provides means of extending and refining the definition based on future research in the field of Complex Software Systems. The properties may be updated or expanded as new information becomes available.
This will require empirical validation of hypotheses and possibly rejecting null hypotheses posed by the definition. The validation is expected to trigger a refinement cycle for the theory, as defined by theory-to-research strategy \cite{swanson2013theory}. The validation may be based on case studies of existing software systems or experimental research based on simulations and modeling.

The list of the initial 13 properties harvested from studies of Complex Systems in other fields may not always have a precise mapping to software systems.
Some properties that define necessary conditions will require further research to understand what they can indicate in the context of software systems. Especially \colorbox{cyan!75}{P7-R} "context-dependency", \colorbox{cyan!75}{P9-R} "path-dependency",  \colorbox{orange!75}{P10-N} "disorder", 
\colorbox{orange!75}{P12-N} "memory" or \colorbox{lime!75}{P13-S} "SoS sufficiency".

The categorization, discussed in Section \ref{categorization}, based on the definition, might provide additional research topics while helping to uncover new underlying general principles of Complex Software Systems.

\section{Conclusions} \label{conclusions}

Complex Systems were identified and studied in different fields, such as physics, biology, and economics.
These systems exhibit properties such as self-organization, robust order, and emergence.
In recent years, software systems started to display behaviors associated with Complex Systems, showing previously unknown potential (e.g., GPT-based applications). However, a commonly shared definition of a Complex Software System is not yet available.

For this reason, in this paper, we have presented a definition of a Complex Software System that can serve as a reference for academia to support future research. The definition is a set of 13 initial \textit{necessary}, \textit{representative}, and \textit{sufficient} conditions for a software system to exhibit Complex System behaviors.
The properties were selected from Complex Systems research in other fields and mapped to software systems. We suggested allowing for evolution and refinement of the properties, as the definition can be refined by evaluating the studied properties using empirical methods. We have also provided examples of the use of the definition and discussed further research directions in the area of Complex Software Systems.

An unambiguous definition of a Complex Software System is a stepping stone toward understanding its underlying principles. 

\section*{Acknowledgement}
\small { The work was supported from ERDF/ESF “CyberSecurity, CyberCrime and Critical Information Infrastructures Center of Excellence” (No. CZ.02.1.01/0.0/0.0/16\_019/0000822).}

\bibliographystyle{myIEEEtran}
\bibliography{ref}

\end{document}